\RequirePackage{lineno}
\documentclass[]{aa}
\usepackage{amsmath} 
\usepackage{amssymb}

\usepackage[utf8]{inputenc}
\usepackage{epsfig}
\usepackage{url}
\usepackage{color}
\definecolor{darkgreen}{rgb}{0,.6,0}
\definecolor{linkcol}{rgb}{.6,0,0}
\usepackage[colorlinks=true,citecolor=darkgreen]{hyperref}

\usepackage{natbib}
\bibpunct{(}{)}{;}{a}{}{,}
\usepackage{subfigure}

\usepackage{upgreek}  

\def\lta{~\raise.4ex\hbox{$<$}\llap{\lower.6ex\hbox{$\sim$}}~}
\def\gta{~\raise.4ex\hbox{$>$}\llap{\lower.6ex\hbox{$\sim$}}~}

\def\msol{\ensuremath{\rm \mass_\odot}} 
\newcommand\solm\msol

\usepackage{dcolumn} \newcolumntype{d}[1]{D{.}{.}{#1}}
\usepackage{multirow}

\bibpunct{(}{)}{;}{a}{}{,}

\begin{document}

\title{In-situ enrichment in heavy elements of hot Jupiters}
\titlerunning{Heavy elements of hot Jupiters}

\author{A. Morbidelli\inst{1}, K. Batygin\inst{2}, E. Lega\inst{1}}


\institute{$^1$ D\'epartement Lagrange, University of Nice -- Sophia Antipolis, CNRS, Observatoire de la C\^{o}te d'Azur, Nice, France; $^2$ GPS Division, Caltech, Pasadena, California}


\date{[Received / accepted]}
 
\abstract
{Radius and mass measurements of short-period giant planets reveal that many of these planets contain a large amount of heavy elements. Although the range of inferred metallicities is broad, planets with more than 100 $M_\oplus$ of heavy elements are not rare. This is in sharp contrast with the expectations of the conventional core-accretion model for the origin of giant planets. } 
{The proposed explanations for the heavy-element enrichment of giant planets  fall short of explaining the most enriched planets. We look for additional processes that can explain the full envelope of inferred enrichments.} 
{We revisit the dynamics of pebbles and dust in the vicinity of giant planets using analytic estimates and published results on the profile of a gap opened by a giant planet, on the radial velocity of the gas with respect to the planet, on the Stokes number of particles in the different parts of the disk and on the consequent dust/gas ratio.  Although our results are derived in the framework of a viscous $\alpha$-disk we also discuss the case of disks driven by angular momentum removal in magnetized winds.} 
{When giant planets are far from the star, dust and pebbles are confined in a pressure bump at the outer edge of the planet-induced gap. Instead, when the planets reach the inner part of the disk ($r_p\ll 2$ au), dust penetrates into the gap together with the gas. The dust/gas ratio can be enhanced by more than an order of magnitude if radial drift of dust is not impeded farther out by other barriers. Thus, hot planets undergoing runaway gas accretion can swallow a large amount of dust, acquiring $\sim 100  M_\oplus$ of heavy elements by the time they reach Jupiter-mass.} 
{Whereas the gas accreted by giant planets in the outer disk is very dust-poor, that accreted by hot planets can be extremely dust-rich. Thus, provided that a large fraction of the atmosphere of hot-Jupiters is accreted in situ, a large amount of dust can be accreted as well. We draw a distinction between this process and pebble accretion (i.e., the capture of dust without the accretion of gas), which is ineffective at small stellocentric radii, even for super-Earths. Giant planets farther out in the disk are extremely effective barriers against the flow of pebbles and dust across their gap. Saturn and Jupiter, after locking into a mutual mean motion resonance and reversing their migration could have accreted small pebble debris.
} 

\maketitle
\section{Introduction}\label{sec:intro}

The discovery and characterization of exoplanets over the course of the last thirty years has brought  a seismic shift in our comprehension of planetary formation and evolution. Nonetheless, the first objects to be discovered in large numbers -- hot-Jupiters -- continue to stand out as an enigmatic class of astrophysical bodies. These giants are close to their host stars (with orbital periods of less than 10 days) and despite their unexpected nature, have attracted extensive scrutiny due to their distinctive features and relative ease of observation. In particular, precise mass and radius determinations are substantially more common within the presently known census of hot-Jupiters than other types of extrasolar planets.

Conventional giant planet structure theory holds that the mass-radius relation for degenerate Jovian planets is approximately flat (i.e., mass-independent: \citet{1982AREPS..10..257S}), meaning that their size is largely dictated by their composition. Within this framework, Jupiter's radius in first approximation corresponds to a roughly solar mixture of hydrogen and helium, meaning that any smaller radius is indicative of a substantially super-solar overall metallicity. The estimate of total mass of heavy elements, here denoted $M_h$, can be sharpened further through detailed modeling (that accounts for the corrections due to age, total mass, etc), and \citet{2006A&A...453L..21G} were the first to carry out this analysis for a pool of 9 well-characterized hot-Jupiters. Intriguingly, they found that some objects have $M_h$ on the order of $\sim 100 M_\oplus$, where $M_\oplus$ denotes the mass of Earth, and pointed out an apparent correlation between the planet's $M_h$ and the metallicity of the central star. This investigation was later extended and confirmed in \citet{2008PhST..130a4023G}, \citet{2011ApJ...729L...7L} and \citet{2013Icar..226.1625M}.

An important complication that arises within such analyses is that hot-Jupiters experience substantial radius inflation, such that their interiors are not in a fully degenerate state. Though a number of physical mechanisms -- including tidal damping \citep{2003ApJ...592..555B,  2007A&A...462L...5L}, breaking gravity waves \citep{2002A&A...385..156G}, impeded cooling due to enhanced atmospheric opacity \citep{2007ApJ...661..502B}, double-diffusive convection \citep{2007ApJ...661L..81C}, turbulent burial of atmospheric entropy \citep{2010ApJ...721.1113Y} and Ohmic dissipation \citep{2010ApJ...714L.238B} -- have been proposed to explain this anomalous heating within these planets' envelopes, statistical analyses \citep{2018AJ....155..214T} have shown that the strong dependence of the degree of inflation on stellar irradiation predicted by the Ohmic dissipation mechanism is indeed reflected in the data (see also cite: \citet{2022A&A...658L...7K}). Consequently, application of conventional giant planet evolution models to strongly-irradiated planets can yield negative heavy element masses. In turn, this implies that the values of $M_h$ reported in the aforementioned studies are lower bounds.

To circumvent this problem, \citet{2016ApJ...831...64T} considered a subset of 47 giant planets that are not strongly irradiated by their central star. Within this subset of objects, anomalous heating of the interior could be reasonably assumed to be negligible, meaning that the computed values of $M_h$ likely represent the actual masses in heavy elements and not their lower bounds (indeed, no negative values of $M_h$ appear in Thorngren et al.'s calculations). We note that, strictly speaking, many of these planets fall into the ``warm Jupiter''  category because they have periods that exceed the nominal 10 day boundary, but for simplicity we still refer to these planets as hot-Jupiters.

\begin{figure}
  \includegraphics[width=10.cm]{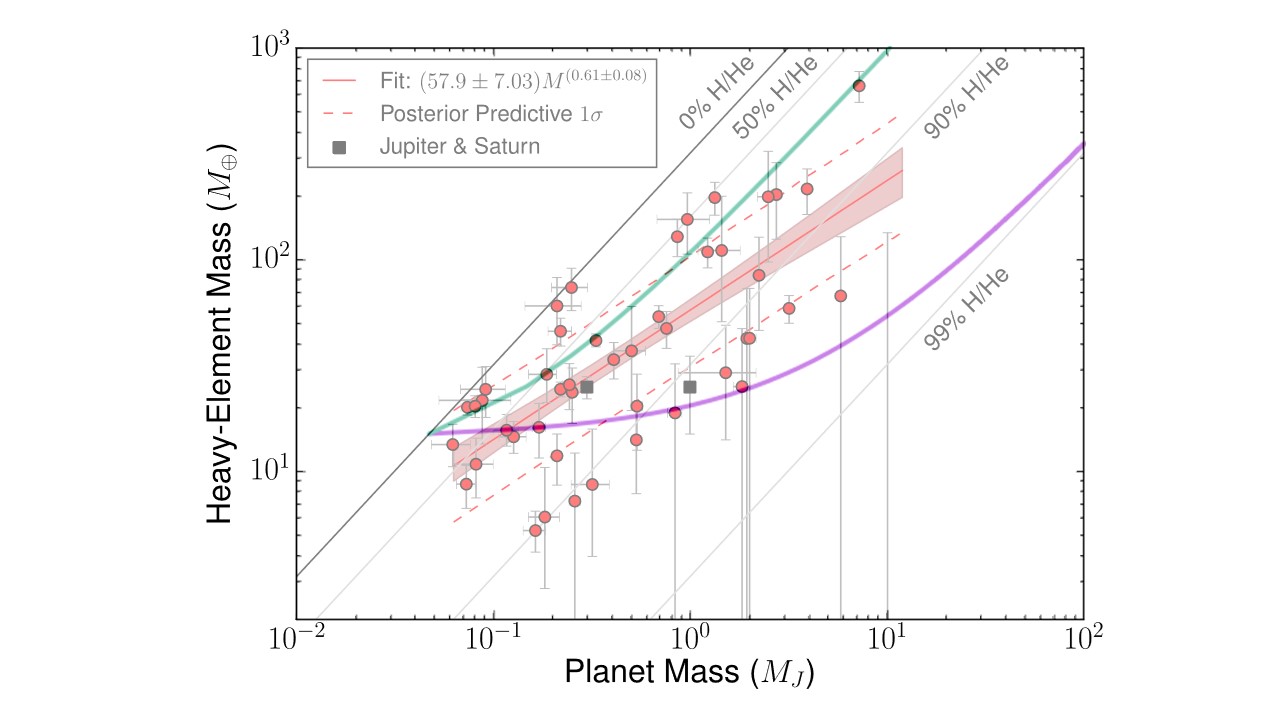}
\vspace*{0.cm}
\caption{A reproduction of Fig. 7 of \citet{2016ApJ...831...64T} showing the mass in heavy elements of weakly irradiated transiting giant planets, as a function of the total planet mass. The relationship (\ref{corr}) is also illustrated. The magenta curve shows the expected enrichment if one considers accretion of gas with 1\% metallicity and of planetesimals following \citet{2020A&A...634A..31V}. { The green curve is for the case with a gas metallicity of 30\%, the most favorable case expected from Sect. ~\ref{alphadisc}}}
\label{Thorn}
\end{figure}

The main result of Thorngren et al. is reproduced in Fig. ~\ref{Thorn} and shows that many hot-Jupiters are more enriched in heavy elements than Jupiter or Saturn. Some Jupiter-mass planets exceed $100$ Earth masses in heavy elements. On average,  the mass in heavy elements $M_h$ is correlated to the total planet mass $M_p$ as: 
\begin{equation}
  M_h = 57.9 \pm 7.0 M_\oplus \left({{M_p}\over{M_{jup}}}\right)^\beta\ ,
  \label{corr}
\end{equation}
where $M_{jup}$ is the mass of Jupiter. The exponent $\beta$ of the $M_h(M_p)$ correlation is $0.61 \pm 0.08$. Thorngren et al. also confirmed the correlation between  $M_h/M_p$ (a.k.a. the planet metallicity) and the stellar metallicity\footnote{The metallicity is usually denoted by the letter $Z$ (and the mass in heavy elements by $M_z$), but we refuse using this notation to protest the Russian invasion of Ukraine, of which $Z$ has become the symbol.}. When this correlation is accounted for, the scatter of the data around the correlation law  (\ref{corr}) is significantly reduced. 

These results are surprising. According to the core-accretion theory of giant planet formation, giant planets are nucleated by the gradual accumulation of solid material into a $\sim 15 M_\oplus$ core, which then accretes a massive envelope of gas and small dust with approximately stellar metallicity. This process is expected to result in a range of solid-to-gas ratios for giant planets that is appreciably super-stellar but nonetheless much smaller than that given by (\ref{corr}). For instance, a Jupiter-mass planet would be expected to have $\sim 18 M_\oplus$ of heavy elements. This mismatch between expectations and observations suggests that the process of hot-Jupiter formation may be more complex than originally thought.

Throngren et al. proposed an explanation for the surprising heavy element enrichment observed in hot-Jupiters. They conjectured that planets accrete all planetesimals located within their feeding zone, an annuls with a radial width proportional to the planet's Hill radius $R_H=a(M_p/3M_{star})^{1/3}$ where $a$ is the semi-major axis of the accreting planet. The exponent 1/3 is smaller than that of (\ref{corr}) but still not grossly inconsistent with the data. \citet{2020A&A...634A..31V} proposed a similar explanation for the enrichment in heavy elements in Jupiter and, through a more sophisticated planetesimal accretion model, predicted an exponent of 2/5 when the formation of a planetesimal gap is considered, i.e. a bit closer to the measured value of  $\beta$ than the estimate of Throngren et al.. The combination of accretion of gas with a 1\% metallicity and the accretion of planetesimals with the $M_p^{2/5}$ relationship gives the magenta curve in Fig.~\ref{Thorn}, which explains some of the planets, but clearly not the majority of them. \citet{2020A&A...633A..33S,2022A&A...659A..28S}  showed that planet migration can enhance the efficiency of planetesimal accretion, due to a combination of resonant shepherding and gas-drag. This is particular efficient as the planet is migrating through specific locations of the disk (dependent on parameters). This may potentially make hot-Jupiters more metal rich than Jupiter itself.  

A radically different mechanism for the heavy element enrichment of giant planets has been proposed by \citet{2021A&A...654A..71S}, elaborating from an original idea of \citet{2006MNRAS.367L..47G}. In their model, inward-drifting dust particles (a.k.a. pebbles) evaporate their volatile elements, each at a specific distance (the sublimation line of the corresponding volatile specie). Therefore, the gas in the inner part of the disk gets enriched in vapor of volatile elements, by a substantial amount for some disk parameters. Because a planet accretes H, He and heavier element vapors indiscriminately, it can be enriched in heavy volatile elements through this process. 

It is worth noting that this model is not restricted to hot-Jupiters, and applies to any planet accreting a substantial fraction of its gaseous envelope in the inner part of the protoplanetary disk, where volatile elements are in vapor state. Nevertheless, this model falls a bit short of reproducing correlation (\ref{corr}). The masses $M_h$ are typically below the value predicted by (\ref{corr}) up to planets with $M_p=2 M_{Jup}$ and can reach the observed mean values only for more massive planets formed in the most metallic disks. Planets with $M_h>100 M_\oplus$ are not expected in the Schneider and Bitsch model unless their total mass is larger than $\sim 3 M_{Jup}$. 

Here we propose that the heavy element enrichment of hot-Jupiters can be explained by the accretion of very dust-polluted gas in the inner part of the disk (essentially in-situ; \citet{2000Icar..143....2B, 2016ApJ...829..114B}). This process is not in contradiction to the picture proposed by Schneider and Bitsch but introduces a previously overlooked effect. It can explain the enrichment in refractory elements, whereas Schneider and Bitsch predict an enrichment only in volatile elements. 

It is generally expected that the gas accreted by a giant planet is metal-poor. This is because dust coagulation converts most of the solid mass into pebble-size objects, which are moderately coupled to the gas; pebble cannot be accreted by planets  exceeding $\sim$ 20 -- 40 $M_\oplus$ (exact value depending on disk's viscosity and scale height; \citet{2014A&A...572A..35L, 2018A&A...612A..30B}) because they remain trapped in the pressure bump produced at the outer edge of the planet-induced gap. 

In this manuscript we will show that, while this is true for giant planets in the central/outer parts of the disk, at small orbital radii pebbles readily permeate the gap of close-in giant planets due to two important factors: (1) the gas flows through the planet's gap in the outside-in direction if the planet is sufficiently close to the star, whereas it flows through the gap in the inside-out direction if the planet is farther out \citep{2015A&A...574A..52D} and (2) pebbles have very small Stokes numbers in the inner part of the disk \citep{2022A&A...666A..19B} and therefore can be entrained into the gap by the radial flow of the gas despite the existence of a pressure bump at the gap's outer edge. Pebbles are expected to fragment while flowing into the gap, thus maintaining a small Stokes number even inside the gap. The strong coupling with the gas then makes the accretion of solids possible only in conjunction with the accretion of gas. Nevertheless, this can increase considerably the planet's heavy element budget because of the very high dust/gas density ratio that can be achieved in the inner disk due to the rapid drop in the dust/gas radial velocity ratio.  

For simplicity, we elaborate on this process in section~\ref{alphadisc}  by adopting a viscous  $\alpha$-disk model, but we then discuss in section \ref{winddisk}  how the process is affected if the radial transport of gas is mostly due to angular momentum removal in disk winds and/or occurs only near the surface of the disk. Thus, we will present a comprehensive revision of the problem of the interaction of a gap-opening planet with the flow of gas and dust or pebbles, hopefully correcting some misconceptions that can be often found in the literature. 

This manuscript will end with a discussion on super-Earths and on the case of Jupiter and Saturn in section \ref{SE-JS}, before the classic summary of the conclusions in section \ref{end}.

\section{Pebble dynamics in the vicinity of a giant planet in an $\alpha$-disk model}
\label{alphadisc}

The dynamics of a pebble in a protoplanetary disk is dictated by its coupling with the gas. Due to gas drag, the pebble's velocity $v$ relative to the gas velocity $u$ is damped following the equation:
\begin{equation}
  {{{\rm d}v}\over{{\rm d}t}} = -{{1}\over{t_f}} (v-u)
\end{equation}
where $t_f$ is called the friction timescale. The Stokes number $S_t$ of a pebble is the value of its friction timescale in units of the local orbital timescale:
\begin{equation}
  S_t= t_f \Omega
\end{equation}
where $\Omega$ is the local orbital frequency.

The radial velocity of a pebble is given by
\begin{equation}
  v_r = {{u_r}\over{1+ S_t^2}} - 2 S_t (v_\theta - u_\theta)\ ,
      \label{vel}
\end{equation}
\citep{1986Icar...67..375N,2002ApJ...581.1344T}, where the $r$ and $\theta$ subscripts denote the radial and azimuthal components of the velocities. { This formula does not account for the back-reaction of dust over gas \citep{2018MNRAS.479.4187D}}.  
For a pebble on a circular orbit, the difference in azimuthal velocities $v_\theta - u_\theta $ is a fraction $\eta$ (depending on $r$) of the Keplerian velocity $v_K$, where
\begin{equation}
  \eta(r) = - {1\over 2}\left({H\over r}\right)^2 {{\partial \log P}\over{\partial \log r}}\ ,
  \label{eta}
\end{equation}
$H$ is the scale height of the disk and $P= (H\Omega)^2 \Sigma /\big(\sqrt{2\pi} H\big)$ is the internal pressure of the gas. Assuming $H\propto r$, i.e. neglecting disk flaring,  (\ref{eta}) can be approximated by
\begin{equation}
  \eta(r) = - {1\over 2} \left({H\over r}\right)^2  \left[{r\over \Sigma} {{\partial \Sigma}\over {\partial r}} -2\right]\ ,
  \label{eta-bis}
\end{equation}
which is the equation we will be using in the rest of this paper.

Eq. (\ref{eta-bis}) reveals that, where $\Sigma$ monotonically decays with $r$, $\eta>0$. In this case, the azimuthal drag of the gas on the pebble leads to the star-ward radial drift of the pebble. However, where $\Sigma$ has a sufficiently positive radial gradient, $\eta<0$ and the drag is reversed. The location where $\eta=0$ is called a pressure bump.

Giant planets open deep gaps in the gas distribution of the disk. Along the outer edge of the gap $\partial \Sigma/\partial r$ is positive and large, so that $\eta<0$, whereas far from the planet's orbit $\eta>0$. So, a pressure bump is established whenever a giant planet forms. If the radial velocity of the gas $u_r > 0$ then pebbles cannot drift into the gap, whatever their Stokes number. If instead $u_r < 0$ only particles with $S_t > \min[u_r(r)/2\eta(r)v_K]$ don't penetrate the gap, where the minimum is computed for $r$ ranging from one to multiple Hill radii beyond the planet's orbit. 

In most hydrodynamical studies on gap opening by giant planets published in the literature, the giant planet is kept on a fixed orbit. The gas is then observed to flow through the gap, from the outer part of the disk to the inner part \citep{1999ApJ...514..344B, 1999MNRAS.303..696K, 1999ApJ...526.1001L, 2018ApJ...854..153W}. Thus, Weber et al. computed that only particles with $S_t>10^{-3}$ do not penetrate a gap opened by a Jupiter-mass planet, for the disk parameters used in their nominal simulations. Consequently, it is often considered in the literature that the so-called {\it planet barrier} against the radial drift of pebbles is effective only for pebbles with a Stokes number larger than this order of magnitude (e.g. \citet{2023A&A...670L...5S}). 

However, the situation is radically different if the planet is allowed to migrate, instead of being kept artificially on a fixed orbit. It has been shown \citep{2014ApJ...792L..10D, 2015A&A...574A..52D, 2018A&A...617A..98R} that, although the migration speed of giant planets is proportional to disk viscosity, giant planets migrate faster than the unperturbed radial velocity of the gas (which, in a viscous disk is $u_r =-3/2 (\nu/r)$, where $\nu$ is the disk's viscosity, expressed as $\nu=\alpha H^2\Omega$ in the so-called $\alpha$-disks) when $\Sigma(r_p)r_p^2/M_p > 0.2$, where $\Sigma(r_p)$ is the gas' unperturbed surface density at the radial distance of the planet $r_p$. In this case, the gas flows through the gap in the inside-out direction (even if the radial motion of the gas in an absolute reference frame can remain negative). Instead, the planet migrates significantly slower than the radial motion of the gas, allowing gas to pass through the gap in the outside-in direction, if
\begin{equation}
  {{\Sigma(r_p)r_p^2}\over {M_p}} << 0.2
\label{pass}
\end{equation}
\citep{2015A&A...574A..52D}. Because of the $r_p^2$ dependence of this formula and the typical weak radial decay of $\Sigma$ (usually proportional to $1/\sqrt{r}$) this happens only when the planet is in the very inner part of the disk. For a typical disk with surface density $175{\,\rm g/cm}^2/\sqrt{(r/5.2\, {\rm au})}$ (delivering a stellar accretion rate of $\sim 4\times 10^{-8} M_\odot/y$ for $\alpha=3\times 10^{-3}$) { and a Jupiter-mass planet,} condition (\ref{pass}) translates to $r_p << 2.5$ au \citep{2018A&A...617A..98R}, which is well satisfied by the planets studied in \citet{2016ApJ...831...64T}.

Obviously, in a reference frame co-moving radially with the planet, $u_r >0$ in the first case (inside-out flow through the gap) and $u_r <0$ in the second case. This means that the planet barrier is effective for particles of {\it all} sizes and Stokes numbers when a giant planet migrates in the central or outer parts of the disk\footnote{only turbulent diffusion in principle can allow some particles to pass through the planet barrier but it has to operate against the gas flow so it is expected to be highly inefficient.}, whereas the barrier starts to be leaky when the planet reaches the vicinity of the star. 

In the following we present some quantitative estimates on the Stokes number of pebbles that can penetrate a gap opened by a giant planet, once the latter is close enough to the star for condition (\ref{pass}) to be true.

\subsection{{ Searching for a dust trap: are particles characterized by a  size or a Stokes number?}}
\label{sec-frag}

Before computing the critical Stokes number below which pebbles can penetrate the gap, we need to determine whether particles { should be characterized by a given size or a given Stokes number, whatever their position in the disk. In fact, if particles are characterized by a given $S_t$, we can apply (\ref{vel}) computing $u_r$ and $\eta$ as function of $r$ and setting $S_t^{crit} = \min[u_r(r)/2\eta(r) v_K]$. If instead particles are characterized by a given size, the Stokes number depends on the gas surface density $\Sigma$ as $S_t= \sqrt{2\pi} {{\rho_p R}\over \Sigma}$, where $\rho_p$ is the bulk density of the pebble of radius $R$; thus, we need to set $S_t(r)=S_t^0 \Sigma(r_0)/\Sigma(r)$, where $S_t^0$ is the Stokes number of the pebble at a reference location $r_0$  and $\Sigma(r)$ is the actual density of the disk due to the presence of the gap, and solve (\ref{vel}) with respect to $S_t^0$.} 

Particles continuously collide with each other, and break or coagulate depending of their collision speed. If the size of particles is limited by the fragmentation barrier, { they achieve at collisional equilibrium a} Stokes number given by \citep{2009A&A...503L...5B}:
\begin{equation}
  S_t^{coll.eq}= v_{frag}^2/(3 \alpha c_s^2)
  \label{Stfrag}
\end{equation}
where $v_{frag}$ is the velocity threshold for fragmentation and $c_s$ is the sound speed. Eq. (\ref{Stfrag}) is independent on the gas surface density $\Sigma$, if not indirectly through $c_s\propto \Sigma^{1/8}$, a weak dependence that we will neglect in the following for simplicity\footnote{$c_s\propto \sqrt{T}$, where $T$ is the gas temperature. The value of the temperature is dictated by the balance between the energy released by accretion of gas towards the star, which is constant through the gap by conservation of mass flow, and cooling, which is proportional to $T^4/(\Sigma f_{dust}\kappa_{dust})$ where $f_{dust}$ is the dust/gas surface density ratio and $\kappa_{dust}$ is the opacity of dust. Thus, $T\propto \Sigma^{1/4}$ and $c_s\propto \Sigma^{1/8}$.}. Eq. (\ref{Stfrag}) applies if the timescale over which a particle experiences a change in $\Sigma$ is longer than the collision timescale with other particles. 

{ In the search for a dust trap at some location along the $\Sigma$-gradient characterizing the gap, it is correct to assume that the particle's Stokes number is that given by (\ref{Stfrag}). In fact, when a particle is trapped, it has the time to experience collisions and grind down until its Stokes decreases to $S_t^{coll.eq}$. Thus, the trapping can be permanent only if the condition $v_r=0$ in (\ref{vel}) holds for $S_t=S_t^{coll.eq}$. Searching for a dust-trap location assuming a fixed dust size is not a valid approximation.}



\subsection{Evaluating the velocity of the gas}

Having determined that $S_t$ is roughly constant and that the critical Stokes number below which particles can penetrate into the gap is $S_t^{crit} = \min[u_r(r)/2\eta(r) v_K]$, we now proceed to evaluate $\eta(r)$ and $u_r(r)$.

We start from the gap profile formula provided in \citet{2006Icar..181..587C}, which gives an expression for $\eta$ as a function of $\Delta=(r-r_p)/R_H, \alpha$ and $q=M_p/M_{star}$:
\begin{equation}
\eta= -{1 \over 2} \left[{{0.4 q^2 r_p^4\left({1\over{R_h\Delta}}\right)^4}\over{{3\over 2}\alpha+{{R_h/r_p}\over{8\Delta}}+200{{R_h/r_p}\over{\Delta^{10}}}}}-2\left({H\over r}\right)^2\right]\ 
  \label{eta-crida}
\end{equation}
where, with respect to formula (14) in Crida et al. we have retained only the term corresponding to (\ref{eta-bis}), assumed $r\sim r_p$ and retained only the dependence on $\Delta$.

Formula (\ref{eta-crida}) gives a minimum of $\eta$ of -0.024 at $\Delta=2.2$ for $q=1\times 10^{-3}$, $\alpha=3\times 10^{-3}$ and $H/r=0.05$, in good agreement with the nominal hydrodynamical simulation of \citet{2018ApJ...854..153W} (-0.03; see their Fig. 2). Instead, for $q=7.6\times 10^{-5}$, $\alpha= 10^{-3}$ and $H/r=0.05$, formula (\ref{eta-crida}) gives $\eta=-0.009$, whereas \citet{2018A&A...612A..30B} found $\eta=0$. It is well known that the model of Crida et al. is not very accurate for planets of moderate mass. Thus, we introduce an empirical correction to (\ref{eta-crida}) by dividing the first term in the $[.]$ by $3.5 (7.6\times 10^{-5}/q + 0.2)$.

Concerning the radial velocity of the gas $u_r$ we remark that, due to the conservation of radial mass flow, one has $u_r=u_r^0 \Sigma^0(r)/\Sigma(r)$, where $u_r^0=-3/2 \alpha (H/r)^2 v_K$ is the gas radial velocity in the unperturbed $\alpha$-disk. Notice that this implicitly assumes that the radial migration of the planet is much slower than $-3/2 \alpha (H/r)^2 v_K$, i.e. that condition (\ref{pass}) holds true. Otherwise we should use the {\it relative velocity} $u_r^0=-3/2 \alpha (H/r)^2 v_K -v_r^P$, where $v_r^P$ is the radial migration velocity of the planet. Again, if $u_r^0>0$ there is no possibility for particles with any $S_t$ to penetrate into the gap. 

To evaluate $\Sigma(r)$, and then $u_r$, we turn again to the analytic gap model of \citet{2006Icar..181..587C}. One has from (\ref{eta-bis}):
\begin{equation}
  \Sigma(r)=\Sigma^0(r)+\int_r^\infty 2 {\Sigma(r) \over r} \left(1-{{\eta(r)}\over{(H/r)^2}}\right) {\rm d}r\ .
  \label{int}
\end{equation}
Eq. (\ref{int}) is implicit, but it can be solved iteratively, starting from $r=\infty$ and setting $\Sigma(\infty)=\Sigma^0(\infty)$. Of course a physical approximation of infinity is 10$R_H$ or so, i.e. well beyond the planet's gap. 

\begin{figure}
  \includegraphics[width=10.cm]{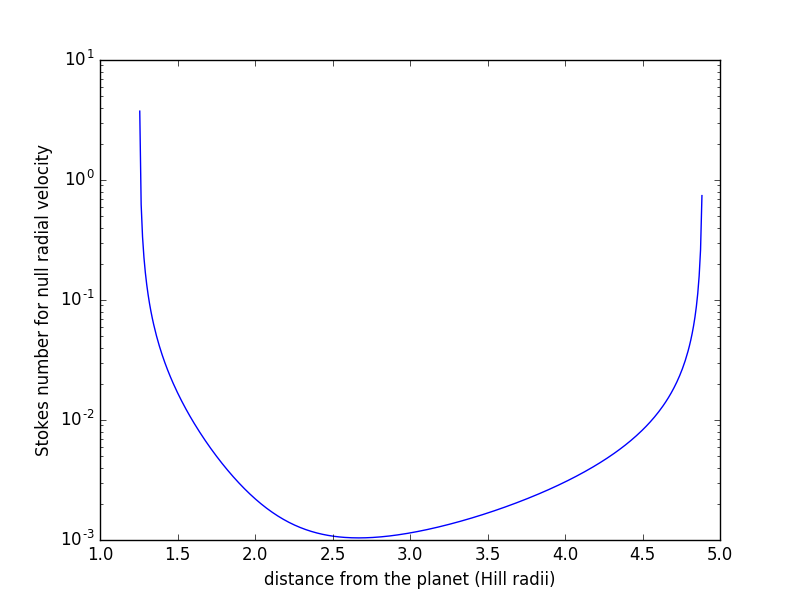}
\vspace*{0.cm}
\caption{The value of the Stokes number of particles with zero radial velocity, as function of the radial distance beyond the planet, normalized to the planet's $R_H$. The parameters used for the calculations are the same as those of the nominal simulation of \citet{2018ApJ...854..153W}: $q=10^{-3}$, $\alpha=3\times 10^{-3}$, $H/r=0.05$. The critical Stokes number is the minimal value along the curve. Compare with the red curve in the bottom panel of Fig. 4 of Weber et al.}
\label{weber}
\end{figure}

Fig.~\ref{weber} shows $S_t(r)$, solution of $v_r=0$ in (\ref{vel}) in the nominal case considered by \citet{2018ApJ...854..153W}. The critical Stokes number $S_t^{crit}$ is the minimum of $S_t(r)$ and is $\sim 10^{-3}$, in good agreement with Weber et al. numerical results. 

\begin{figure}
  \includegraphics[width=10.cm]{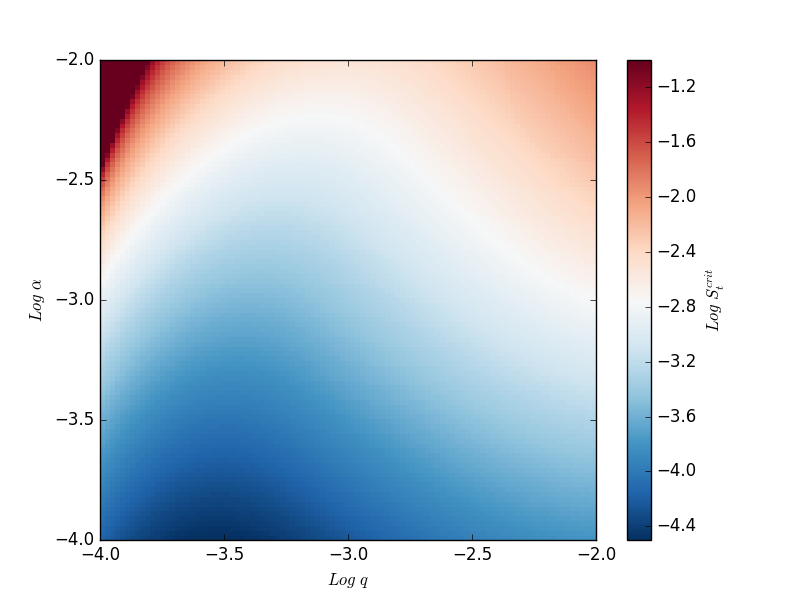}
  \includegraphics[width=10.cm]{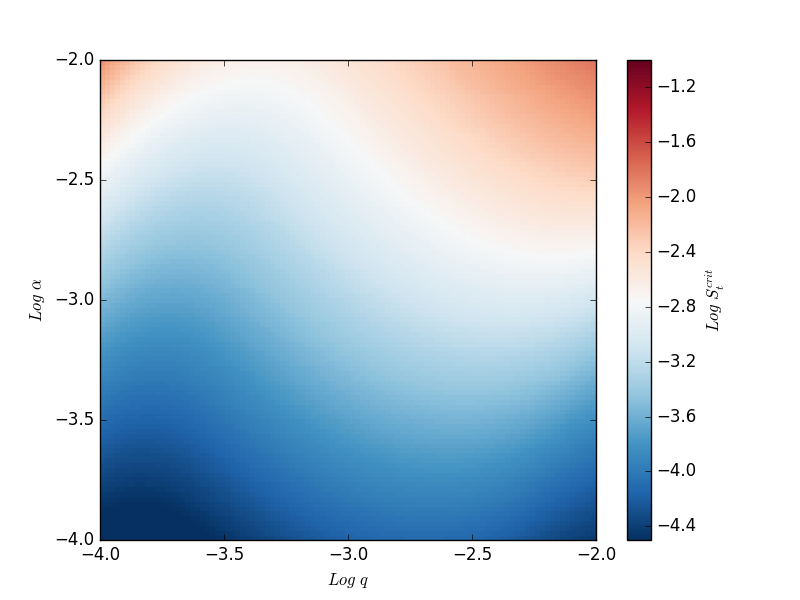}
\vspace*{0.cm}
\caption{The critical Stokes number as a function of planetary mass $q$ (normalized to the stellar mass) and viscosity parameter $\alpha$, for $H/r=0.05$ (top) and $H/r=0.03$ (bottom). The dark red color in the top left corner of the top panel denotes a region of parameter space where particles with any Stokes number penetrate into the gap.}
\label{St-map}
\end{figure}

Fig.~\ref{St-map} shows a map of the critical Stokes number as a function of $q$ and $\alpha$ for $H/r=5$\% (top) and 3\% (bottom). We remark that the two panels are quite similar, revealing a weak dependence on $H/r$. It is also worth noticing that, for a given value of $\alpha$, the critical Stokes number does not monotonically decrease with increasing planet's mass, as one could naively expect. This is due to the fact that, although the pressure bump becomes stronger with increasing $q$, $-u_r(r)$ increases at any value of $r$ because the gap becomes wider and deeper. Consequently, the value $S_t^{crit}$ is achieved farther away from the planet and can result bigger than that computed for a smaller planet. According to Fig.~\ref{St-map}, for $H/r=0.05$ the planets that oppose the most severe barriers to pebble drift are those of approximately Saturn's mass ($\log q \sim -3.5$).   

\begin{figure}
\includegraphics[width=8.5cm]{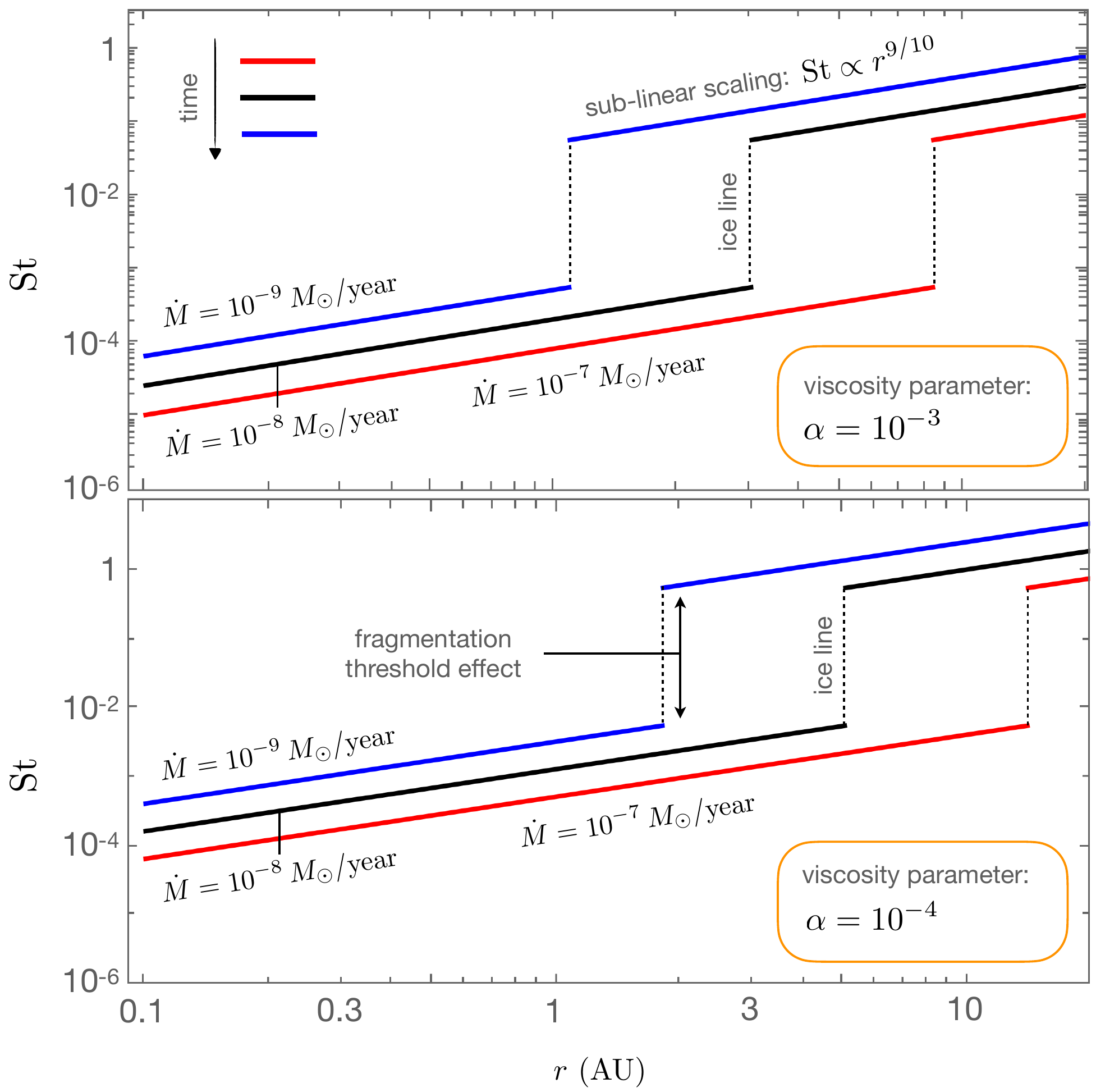}\vspace*{0.cm}
\caption{A re-edition of Fig. of \citet{2022A&A...666A..19B}, recentered in the region of interest, from 0.01 to 10 au.}
\label{BM22}
\end{figure}

It is clear from Fig.~\ref{St-map} that, even for the least viscous  ($\alpha\sim 10^{-4}$) and shallow ($H/r=0.03$) disks, giant planets' gaps are leaky for pebbles with $S_t\lesssim 10^{-4}$. This value is small, but it is the characteristic Stokes number of pebbles that reach the fragmentation limit (\ref{Stfrag}) in the inner part of the protoplanetary disk, where condition (\ref{pass}) is fulfilled (Fig.~\ref{BM22}). In particular, when planets approach the very inner part of the disk where the magneto-rotational instability (MRI) is active, the $\alpha$ parameter is expected to increase from $10^{-4}$ towards several times $10^{-3}$ \citep{2016ApJ...827..144F}. If $\alpha\sim 10^{-3}$, Fig.~\ref{BM22}  shows that pebbles have $S_t<10^{-4}$, which is significantly smaller than $S_t^{crit}$ (Fig.~\ref{St-map}).

Thus, we conclude that when giant planets migrate near the central star, eventually pebble isolation disappears and planets can potentially feed from the full radial flow of solid (refractory) material. The efficiency of the accretion process is discussed next.

\subsection{Once in the gap: accretion on the planet}

{ Particles with $S_t<S_t^{crit}$ penetrate into the gap. For these particles the evolution of the Stokes number is not obvious. In fact, it is important to remember that Eq. (\ref{Stfrag}) only applies if the timescale over which a particle experiences a change in $\Sigma$ is longer than the collision timescale with other particles.} Because the particles radial speed increases as they go into the gap as $\Sigma^0(r)/\Sigma(r)$, where $\Sigma^0(r)$ is the unperturbed surface density and $\Sigma(r)$ the actual density, it is possible that the migration timescale becomes shorter, so that particles preserve their sizes and increase in Stokes number. By comparison of the timescale of unperturbed radial motion of particles coupled to the gas, i.e $T_{drift}=r/u_r=2/[3 \alpha (H/r)^2 \Omega]$, with the particle collision timescale $T_{coll}=1/[f_{dust}\Omega]$ and adopting nominal parameters ($\alpha=10^{-3}$, $(H/r)=0.05$ and $f_{dust}=0.01$), we find that drift timescale becomes shorter than the collision timescale only in gaps that are at least three orders of magnitude deep, namely for planets significantly more massive than Jupiter.

If the gas is very dust-rich, as argued below (Sect.~\ref{dustydisk}), the particle collision timescale becomes even shorter. However, for large values of $f_{dust}$ the stirring of dust by turbulent diffusion in the gas is reduced by their collective inertia (see  \citet{2019ApJ...886L..36S} for laboratory experiments and Sect.~4 of \citet{2020ApJ...894..143B} for an analytic derivation). This lowers the impact velocity and raises the value of the Stokes number at the fragmentation threshold.

Nevertheless we consider it unlikely that the Stokes number of particles can increase by orders of magnitude. Thus, particles should remain very coupled to the gas (e.g. $S_t \sim 10^{-4}$ -- $10^{-3}$). The usual formul\ae\ for pebble accretion cannot be applied in our case because they are derived for small planets that do not accrete (nor perturb) the surrounding gas. Instead, the dynamics of gas in the vicinity of a giant planet is very perturbed. Hydrodynamical simulations (e.g. \citet{2019A&A...630A..82L}) show that part of the gas entering the planet's Hill sphere is accreted into a bound atmosphere, while some merely passes through the Hill sphere with a residence timescale typically shorter than the keplerian orbital period around the planet itself. Given the small Stokes number, particles coming with the accreted flow of gas will also be accreted in the envelope, whereas those carried by the unbound flow will not have enough time to decouple from the background flow and will be eventually transported away. Although solid particles can be accreted in this regime, they do so along with the gas. For this reason, it would be misleading refer to this process as pebble-accretion, since this term specifically refers to the selective accretion of dust over gas.

The accretion of dust together with gas may suggest that the overall planet metallicity cannot increase in this process. However, we show below that the gas in the inner part of a planetary disk can be very dust-rich, with $f_{dust}$ that can approach unity.

\subsection{Metallicity of the inner part of the disk}
\label{dustydisk}

In steady state, if dust drifts radially in the disk at the speed $v_r$ and gas at the speed $u_r$, the dust/gas ratio $f_{dust}$ is given by:
\begin{equation}
  f_{dust}(r)=f_{dust}(r_0) {u_r \over v_r}(r) {v_r \over u_r}(r_0)
  \label{fdust}
\end{equation}
where $r_0$ is a reference distance in the disk. Correspondingly, from (\ref{vel}) we have:
\begin{equation}
  {v_r \over u_r} = 1 + 2 S_t \eta {v_K \over u_r} =  1 + {4\over 3} S_t \eta \alpha^{-1} \left({H \over r}\right)^{-2}
\label{voveru}
\end{equation}
where for $u_r$ we have assumed the usual formula for an unflared $\alpha$-disk.

To fix ideas, we assume $H/r=0.05$ and $\eta=0.003$ \citep{2015A&A...575A..28B}. 
Following \citet{2016ApJ...827..144F}, we also assume
\begin{equation}
\alpha(r)= {{10^{-2}-10^{-4}}\over 2} \left[1-\tanh\left(10-{1\over r}\right)\right]+10^{-4}\ ,
\label{alphaFlock}
\end{equation}
where we have adopted a $1/r$ dependence of the temperature and a critical temperature of $T_{MRI}\sim 1,000$~K to activate the MRI at 0.1~au. Finally, for $S_t$ we use the value derived from a self-consistent description of an $\alpha$-disk in \citet{2022A&A...666A..19B}:
\begin{equation}
  S_t(r)=10^{-4} r^{9/10} \left({10^{-3}\over \alpha}\right)^{4/5}\ .
  \label{StBM}
\end{equation}

\begin{figure}
  \includegraphics[width=10.cm]{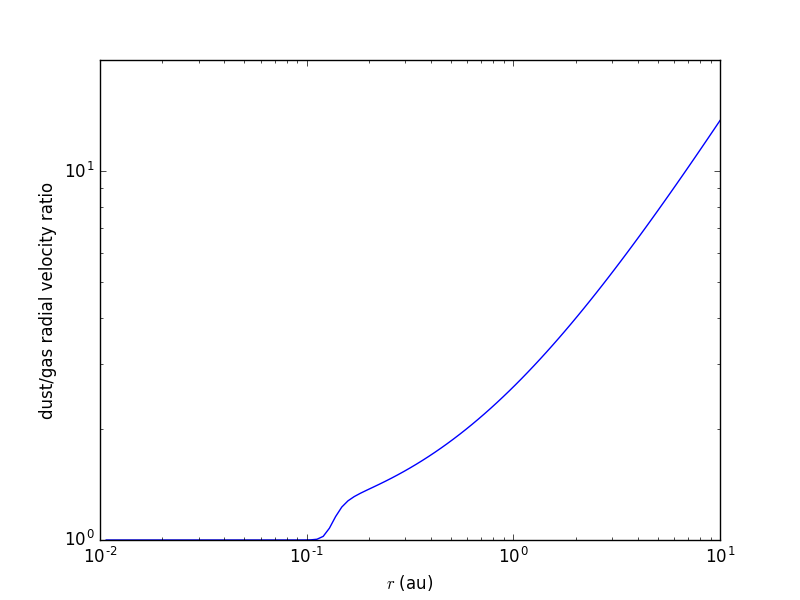}
\vspace*{0.cm}
\caption{The dust/gas radial velocity ratio $v_r/u_r$ as given by (\ref{voveru}), as a function of radial distance from the star. The validity of the curve extends to the location of the snowline (2 to 15 au, depending on the stellar accretion rate for the assumed $\alpha=10^{-4}$, see \citet{2022A&A...666A..19B}), where $S_t$ (and hence  $v_r/u_r$) jumps up by a factor equal to the square of the ratio of fragmentation velocity thresholds for icy and rocky particles, respectively.}
\label{Figvu}
\end{figure}

Using (\ref{alphaFlock}) and (\ref{StBM}), the resulting dust/gas radial velocity ratio (\ref{voveru}) is illustrated in Fig.~\ref{Figvu}. Recall from (\ref{fdust}) that the dust/gas ratio, a.k.a. gas metalliticy, is inversely proportional to  (\ref{voveru}). Thus, in the inner disk, the metallicity can be $\sim 2.5$ times higher than at  1 au and $\sim 6$ times higher than at 3 au.

To use this result, we need to anchor the dust/gas ratio somewhere in the disk. In a steady-state scenario, where the dust radial velocity is dominated by the $-2 S_t \eta v_K$ term in (\ref{vel}) the dust to gas ratio is (\citet{2016A&A...591A..72I}, formula 46):
\begin{equation}
  f_{dust} = 2\times 10^{-3} \left({S_t \over 0.1}\right)^{-1} \left({t \over {\rm 1 My}}\right)^{-1/3}\ ,
    \label{Ida}
\end{equation}
where $t$ is the age of the disk. We recognize the inverse dependence on $S_t$ (itself proportional to $r^{9/10}$ -see eq. \ref{StBM}) beyond $\sim 0.3$~au in Fig.~\ref{Figvu}. Formula (\ref{Ida}) is valid only until the {\it pebble formation front} $r_{pf}\sim 50{\rm au} (t/{\rm 1 My})^{7/3}$ \citep{2016A&A...591A..72I} reaches the outer edge of the disk. After this event, the dust flux drops and so drops also the dust/gas ratio.  Even before that time, there may be obstacles to dust drift, due to gaps opened by distant giant planets \citep{2018ApJ...854..153W} or the formation of dust-trapping rings due to non-ideal MHD effects \citep{2020A&A...639A..95R}. Until all these limitations become real, formula \ref{Ida} predicts $f_{dust}=0.2$ at 1~au, where $S_t \sim 10^{-3}$, and $t\sim 1$~My, i.e. enhanced by a factor 50 with respect to solar metallicity. However, the enhancement has to be limited by planetesimal formation. For $S_t\gtrsim 10^{-3}$ the streaming instability converts dust into planetesimals when the dust/gas volume density ratio on the midplane is of order unity \citep{2021ApJ...919..107L}. Depending on the disk viscosity and its ability to stir the vertical distribution of dust (usually encapsulated in the so-called Schmidt number $S_c$) the vertical scale height of the dust layer can be $\sim 1/10$ to $\sim 1/3$ that of the gas disk (i.e. for $\alpha=10^{-4}$ and $S_c=10$ ans $1$ respectively); consequently, when $f_{dust}$ exceeds $\sim 0.1$--$0.3$ planetesimal formation is expected to convert the dust excess into macroscopic bodies. For this reason, we assume that $f_{dust}$ at 1 au cannot exceed these values. Therefore we predict that, in the best case scenario, $f_{dust}$ in the region of warm and hot-Jupiters, where planetesimal formation is inhibited by the Stokes number being $\ll 10^{-3}$, can be 0.3 to 1 respectively, i.e. enriched in metallicity by a factor 30 to 100. { The dust/gas ratio remains nevertheless smaller than one, which justfies the use of Eq. (\ref{vel}) for the dust velocity, even if it neglects the back-reaction of dust on gas. }

The green curve in Fig.~\ref{Thorn} shows the mass in heavy elements expected for giant planet accreting most of their atmosphere in the inner disk, from a gas enhanced in metallicity by a factor 30. { Notice that we don't expect all planets to lay on this curve; the curve illustrates  how metal-rich hot-Jupiters can become under the most favorable conditions}. Correspondingly, a large scatter of values are expected, depending on the actual metallicity of gas and the fraction of the planet's atmosphere accreted in the inner disk.  

\section{Wind-driven disks and layered accretion}
\label{winddisk}

One-dimensional $\alpha$-disk models -- of the type we have employed thus far -- are routinely adopted for their simplicity and the ability to obtain analytical estimates. Their realism, however, is diminished in part by the fact that they do not account for other modes of angular momentum transfer within the system. In this section, we address this drawback by describing what we expect to happen if the flow of gas towards the central star ensues due to angular momentum removal in disk winds, and if the transport of gas occurs predominantly on the surface layer of the disk. This is typical of disks where Ohmic diffusion dominates \citep{2021A&A...650A..35L}, i.e. in the inner disk region ($0.1 < r < 1$~au, \citet{2011ARA&A..49..195A}), but can also occur in viscous disks, due to the meridional circulation unveiled in \citet{2002ApJ...581.1344T} or if there is a deadzone near the midplane \citep{2014A&A...570A..75B}.

Let's start with wind-driven disks. A main difference with the $\alpha$-disk considered in the previous section is the radial dependence of the gas radial velocity $u_r$. As we have seen above, in an $\alpha$-disk $u_r$ scales with $v_K$. Thus, the balance between radial drag and pressure bump (see eq. \ref{vel}) is independent of the distance from the star and only depends on the particles' Stokes number. Particles penetrate into the gap in the inner part of the disk because their Stokes number is smaller there. In a wind-driven disk, instead, $u_r$ scales as $1/r$, i.e. increases faster than $v_K$ as the gas approaches the star \citep{2022A&A...658A..32L}. This is because the wind removes angular momentum only from a sufficiently ionized layer and ionization depends on the amount of gas encountered by radiation as it penetrates from the surface of the disk towards the mid-plane. Thus, the vertically-integrated column density of gas of the ``active'' layer of the disk, where the gas flows towards the central star, is independent of $r$. Then, conservation of mass-flow implies $u_r\propto 1/r$. In this case, it is even easier for particles to penetrate into a planet-carved gap in the inner disk, because the radial entrainment due to the gas radial velocity is stronger. For the same reason, the condition for outside-in flow of gas across the planet gap (a necessary condition for particles of any size to reach the planet) is no longer given by (\ref{pass}) and in general can be fulfilled farther out in the disk \citep{2022A&A...658A..32L}. To this end, \citet{2023ApJ...946....5A} reported an acceleration of the gas radial velocity near the gap due to a concentration of magnetic field lines there. In other words, the scenario envisioned above appears even more favorable for efficient accretion of dust by hot-Jupiters if the disk's evolution is driven by magnetized winds.

The fact that the flow of gas towards the star occurs near the surface in wind-driven disks (and also in some 2D $\alpha$-disks models, with or without dead zone) is not an obstacle to dust accretion. The small Stokes number of particles previously considered ($S_t\sim 10^{-4}$) ensures that the dust is well coupled to the gas and uniformly distributed in the vertical direction of the disk, unless the turbulent vertical stirring is pathologically low ($\alpha/S_c \ll 10^{-4}$), which is unlikely, particularly in the inner disk. Under these conditions, the flow of dust from a given disk layer into the gap prompts the vertical redistribution of the dust. This prevents the dust from accumulating indefinitely on the midplane, even if the pressure bump operates there.

This argument can be formalized. To fix ideas, let's imagine a disk of gas where $u_r=0$ everywhere but in a near--surface layer, where $u_r$ is large enough to transport the dust in that layer into the gap. The vertically integrated radial mass-flow of dust towards the pressure bump is
\begin{equation}
  \dot{M}_r= 4\pi r \eta v_K S_t \Sigma_d \ ,
  \label{flow-r}
\end{equation}
where $\eta\sim 3\times 10^{-3}$ is the value of $\eta$ beyond the pressure bump and $\Sigma_d=f_{dust}\Sigma_g$ is the surface density of the dust. The vertical flow of dust to restore a vertically uniform distribution of the dust/gas ratio is
\begin{equation}
  \dot{M}_z= 2 \pi r \Delta r D \rho_g {\partial \over{\partial z}}\left({\rho_d\over \rho_g}\right) \ ,
  \label{flow-z}
\end{equation}
where $\rho_d(z)$ and $\rho_g(z)$ are the volume densities of dust and gas at height $z$ and $\Delta r$ is the width of the ring where the dust tends to be concentrated due to the pressure bump on the mid-plane. The latter is $\Delta r= r\Delta w_0 \sqrt{\alpha/S_t}$ \citep{2018ApJ...869L..46D}, where $\Delta w_0$ is the radial width of the pressure bump in normalized units. In (\ref{flow-z}) $D=\alpha H^2 \Omega$ is the diffusion coefficient. Because on the surface layer of the disk the dust flows into the gap, $\rho_d$ is reset to $f_{dust}\rho_g$ there and therefore we can approximate ${\partial/{\partial z}}\left({\rho_d/ \rho_g}\right)\sim 1/H [(\rho_d/\rho_g)_{z=0}-f_{dust}]$.

In a steady state (\ref{flow-r}) and (\ref{flow-z}) have to be equal. Approximating $\rho_g\sim \Sigma_g/(2H)$, this gives:
\begin{equation}
  \left({\rho_d\over \rho_g}\right)_{z=0}= \left[ 1+4 \left({S_t\over \alpha}\right)^{3/2} {{\eta}\over {\Delta w_0}}\right] f_{dust} \ .
  \label{dustgasmid}
\end{equation}
The term $S_t/\alpha$ is of order unity, as required for a uniform vertical dust distribution. The ratio $\eta/\Delta w_0$ is typically much smaller than 1/4, $\Delta w_0$ being $\sim 0.1$ for a pressure bump induced by a Jupiter-mass planet \citep{2018ApJ...854..153W}. Thus, the dust/gas ratio on the midplane at the pressure bump is only moderately increased, by a factor $(1+4 \eta/\Delta w_0)<2$ with respect to the local, vertically integrated disk metallicity $f_{dust}$. Once this moderate enrichment is achieved, a steady state dust flux is set and all the net dust flow is carried into the gap, preventing any further accumulation of dust at the pressure bump in the midplane. In particular, it is unlikely that planetesimals would start to form in the mid-plane near a giant planet gap if they could not form in absence of the planet. 

\section{The case of super-Earths and of Jupiter and Saturn in the solar system}
\label{SE-JS}

If giant planets cannot block the flux of dust in the inner part of the disk, the case is considerably more hopeless for super-Earths, as they open much shallower gaps. This does {\it not} imply that close-in super-Earth grew efficiently in situ by pebble accretion. The small Stokes number of the dust and their uniform vertical distribution in the disk makes pebble accretion an inefficient 3D process \citep{2022A&A...666A..19B}. Indeed, as we have seen above, even for giant planets the accretion of dust has to occur together with the accretion of gas. Given that super-Earths, by definition, accreted only moderate quantities of gas, we don't expect that the accretion of dust delivered a substantial fraction of a super-Earth's solid mass. { Planets} with $M_p\sim M_h\sim 70 M_\oplus$, as visible in Fig.~\ref{Thorn}, are not reproduced in our model starting from a 15~$M_\oplus$ core, as the green curve in the figure shows. Indeed, within the framework of our picture, these objects require the accretion of a large amount of planetesimals or mutual merging of multiple super-Earths of smaller masses.

Jupiter and Saturn are also enriched in heavy elements relative to solar metallicity. As individual planets in the outer disk, condition (\ref{pass}) would not be satisfied because the gas would flow through their gaps in the inside-out direction. However, Jupiter and Saturn have the tendency to lock in a mean motion resonance within the nebula, which halts or reverses their migration direction \citep{2001MNRAS.320L..55M, 2007Icar..191..158M, 2023arXiv230304652G}. Once this happens, the gas flows in the outside-in direction through their common gap, carrying small-enough dust with it. The typical Stokes number of particles at $\sim 5$~au should be much larger than the threshold of $10^{-4}$-- $10^{-3}$ for transport into the gap (Fig.~\ref{St-map} and \citet{2018ApJ...854..153W}), but particle fragmentation at the pressure bump can produce a small-end tail in the particle size distribution with Stokes numbers smaller than this threshold \citep{2023A&A...670L...5S}. These particles would be accreted by the planets with the same efficiency as gas (i.e. up to 90\%; \citet{1999ApJ...526.1001L})\footnote{This implies that only a minority of these small particles reached the inner part of the disk, unlikely to contaminate the so-called solar system isotopic dichotomy \citep{2017PNAS..114.6712K}.}. However, it is unlikely that the presence of these particles would have made the metallicity of the accreted gas super-solar given that, as fragments, they don't represent the bulk of the solid mass and the outer disk metallicity is not expected to be substantially enriched, unlike in the inner disk (Sect.~\ref{dustydisk}). Thus we conclude that the enrichment in heavy elements of Jupiter and Saturn is likely due to accretion of planetesimals \citep{2020A&A...634A..31V} and volatile-element vapors \citep{2006MNRAS.367L..47G}.  

\section{Conclusions}
\label{end}

In this work we have analyzed the dynamics of particles near the outer edge of a gap opened by a giant planet in the gas radial distribution. The edge of a gap is a pressure bump but it also enhances the radial velocity of the gas. Giant planet Type-II migration towards the star is typically faster than the radial velocity of gas in the outer part of the disk, and slower in the inner part. Thus, in the outer disk, giant planets are effective barriers against the flow of dust of any size, because both the positive radial motion of the gas relative to the planet and the pressure bump prevent particle drift into the gap. Instead, in the inner part of the disk the radial flow of gas relative to the planet is in the outside-in direction and can entrain particles with small Stokes number into the gap. We find that for the radial entrainement to ensue despite the existence of a pressure bump, the particles' Stokes number has to be smaller than $10^{-4}$--$10^{-3}$, depending on planet mass, disk viscosity and scale height. For particles whose size is limited by the velocity fragmentation threshold, the Stokes number scales inversely with the square of the gas sound-speed and therefore it decreases rapidly in the inner disk. Thus, in the inner disk all conditions are met for typical particles to flow into the gaps opened by giant planets.

As an aside, this implies that the so-called inside-out planet formation model \citep{2014ApJ...780...53C} is unlikely to be operational. In fact that model relies on the ability of the first planet (either formed at or migrated to the inner edge of the disk) to create a pressure bump where drifting dust particles accumulate until forming a second planet and so on. If even a giant planet is not able to block the drift of particles in the innermost part of the disk, this process should not promote the formation of super-Earth systems.  

Returning to giant planet metal-enrichment, we showed that particles, once in the gap, can be accreted by a planet only together with the gas, because of the smallness of the Stokes number. This does not, however, mean that the accreted material has stellar metallicity. In fact, if the dust is allowed to freely drift towards the star (i.e. it is not blocked farther out by a dynamical barrier), it naturally piles up in the inner disk enhancing the local metallicity by an order of magnitude or more. For this reason, if hot-Jupiters accreted a substantial fraction of their envelope in-situ { (see \citet{2000Icar..143....2B, 2016ApJ...829..114B}, but these works should be actualized to account for the high opacity of the dust-rich gas demonstrated in this paper)}, the large enrichments in heavy elements deduced from measurements of the planet's mass and radius could have been achieved by this process.

Although we have based our analysis on an $\alpha$-disk model, we showed that it also holds if the flow of gas in the disk is dominated by angular momentum removal in magnetized winds. We also showed that, because of the small Stokes number, the transport of gas and dust into the gap in a surface layer of the disk is sufficient to prevent dust pile-up at the outer edge of the gap and ensures that, in a steady state, the full dust flux crosses the orbit of the planet. To sum up, this study has explored the multifaceted relationship between particles, gas flow and giant planet migration near gap edges, contributing to a better understanding of the conditions under which particles can enter these gaps. Our findings may help inform future research on the heavy element enrichment observed in hot-Jupiters.

\section{Acknowledgments}
A.M. is grateful for support from the ERC advanced grant HolyEarth N. 101019380 and to Caltech for the visiting professor program that he could benefit from. K. B. is grateful to Caltech's center for comparative planetology, the David and Lucile Packard Foundation, and the National Science Foundation (grant number: AST 2109276) for their generous support.

\end{document}